\documentclass[namedreferences]{SolarPhysics}
\usepackage[optionalrh]{spr-sola-addons} % For Solar Physics
\usepackage{epsfig}          % For epss, old commands
\usepackage{graphicx}        % For eps figures, newer & more powerfull
\usepackage{amssymb}        % useful mathematical symbols
\usepackage{color}           % For color text: \color command
\usepackage{url}             % For breaking URLs easily trough lines
\newcommand{\aap}{    {\it Astron. Astrophys.}}

\newcommand{\aapr}{   {\it Astron. Astrophys. Rev.}}

\newcommand{\apj}{    {\it Astrophys. J.}}
\newcommand{\apjl}{   {\it Astrophys. J.}}
\newcommand{\apjs}{   {\it Astrophys. J. Sup. Ser.}}

\newcommand{\cjaa}{   {\it Chinese J. Astron. Astrophys.}}

\newcommand{\grl}{    {\it Geophys. Res.}}

\newcommand{\jgr}{    {\it J. Geophys. Res.}}

\newcommand{\nat}{    {\it Nature}}
\newcommand{\nar}{    {\it New Astron. Rev.}}

\newcommand{\solphys}{{\it Solar Phys.}}

\newcommand{\ssr}{    {\it Space Sci. Rev.}}

\begin{document}
\begin{article}
\begin{opening}

%%%%%%%%%%%%%%%%%%%%%%%%%%%%%%%%%%%%%%%%%%%%%%%%%%%%%%%%%%%%%%%%%%%%%%%%%
%               Title Page                                              %
%%%%%%%%%%%%%%%%%%%%%%%%%%%%%%%%%%%%%%%%%%%%%%%%%%%%%%%%%%%%%%%%%%%%%%%%
%

\title{Filament Eruption in NOAA 11093 Leading to a Two-Ribbon M1.0 Class Flare and CME}

\author{P.Vemareddy$^{1}$\sep
        R.A.Maurya$^{2}$\sep
        A.Ambastha$^{3}$
         }
\runningauthor{P.Vemareddy et al.}
\runningtitle{Filament Eruption Leading to Flare/CME}

\institute{{Udaipur Solar Observatory (Physical Research Laboratory),
\\ P.O. Box 198, Dewali, Badi Road, Udaipur 313 001, India.}\\
              $^{1}$ e.mail: \url{vema@prl.res.in}\\
              $^{2}$ e.mail: \url{ramajor@prl.res.in}\\
              $^{3}$ e.mail: \url{ambastha@prl.res.in}\\
             }

%%%%%%%%%%%%%%%%%%%%%%%%%%%%%%%%%%%%%%%%%%%%%%%%%%%%%%%%%%%%%%
%                    ABSTRACT                                %
%%%%%%%%%%%%%%%%%%%%%%%%%%%%%%%%%%%%%%%%%%%%%%%%%%%%%%%%%%%%%%
\begin{abstract}
We present multi-wavelength analysis of an eruption event that
occurred in Active Region (AR) NOAA 11093 on 7 August 2010, using
data obtained from SDO, STEREO, RHESSI and GONG H$\alpha$ network
telescope. From these observations, we inferred that upward slow
rising motion of an inverse S-shaped filament lying along the
polarity inversion line (PIL) resulted in a CME subsequent to a
two-ribbon flare. Interaction of overlying field line across the
filament with side lobe field lines, associated EUV brightening, and
flux emergence/cancellation around the filament were the
observational signatures of the processes leading to its
destabilization and the onset of eruption. Moreover, the rising
motion profile of filament/flux rope  corresponded well  with flare
characteristics, viz., the reconnection rate and HXR emission
profiles. Flux rope accelerated to the maximum velocities % as a CME at the peak phase of the flare,
followed by deceleration to an average velocity of 590 kms$^{-1}$.
We suggest that the observed emergence/cancellation of magnetic
fluxes near the filament caused it to rise, resulting in the tethers
to cut and reconnection to take place beneath the filament; in
agreement with the tether cutting model. The corresponding
increase/decrease in positive/negative photospheric fluxes found in
the post-peak phase of the eruption  provides unambiguous evidence
of reconnection as a consequence of tether cutting.

\end{abstract}
\end{opening}

%%%%%%%%%%%%%%%%%%%%%%%%%%%%%%%%%%%%%%%%%%%%%%%%%%%%%%%%%%%
%                    INTRODUCTION                         %
%%%%%%%%%%%%%%%%%%%%%%%%%%%%%%%%%%%%%%%%%%%%%%%%%%%%%%%%%%%

\section{Introduction}
\label{Intro}

Solar filament eruptions are energetic events occurring due to the
explosive release of magnetic energy. Understanding the driver and
trigger mechanisms of these eruptions is one of the most
challenging, ongoing research problems in solar physics. These
events manifest as prominence eruptions if seen at the limb,  and as
filament eruptions if seen against the disk. They can be broadly
divided into two classes - (i) ejective eruptions which give CMEs
and long duration two-ribbon flares, and (ii) confined eruptions
which give short duration flares. However, there are also
filament/prominence eruptions that are not associated with any
flares. What defines the eruption to be ejective or confined event
remains an unanswered question (For detailed reviews on theories of
eruptions, refer to \inlinecite{priest2002}, \inlinecite{lin2003}).

In filament eruptions, one commonly observed feature in
chromospheric H$\alpha$ images is two bright flare ribbons
separating away as the flare progresses. This process is explained
as follows: Reconnection of magnetic fields is believed to be the
underlying mechanism for energy release as proposed by
\inlinecite{carmi1964}, \inlinecite{sturrock1966},
\inlinecite{hirayama1974}, \inlinecite{kopp1976}. As a coronal
magnetic flux rope loses equilibrium and travels upwards, an extreme
reconnection current sheet (RCS) is formed underneath. Reconnection
in this RCS releases most of the magnetic energy stored in the
magnetic field configuration \cite{forbes1984,lin2000}. Charged
particles can be effectively accelerated by the electric field in
the RCS \cite{martens1990,litv1995}.  Some particles, energized
during a solar flare, gyrate around the field lines and propagate
toward the underlying foot points, precipitating at different layers
of the solar atmosphere to produce the two-ribbon flares. Separation
of these chromospheric flare ribbons is believed to provide a
signature of the reconnection process occurring progressively higher
up in the corona. This is the  standard flare model scenario. All
eruption models lead to this standard model in the onset phase of
eruption.

Several ideas have been advanced for explaining the eruption  onset
mechanism \cite{klim2001,forbes2006,moore2006}. In tether-cutting
model proposed by \inlinecite{moore1980}, and further elaborated by
\inlinecite{moore2001}, magnetic tension restraining the sheared
core field of a bipolar magnetic arcade is released by internal
reconnection above the PIL. Evidence for the tether-cutting model
can be found in several recent observational studies
\cite{liu2007,wang2006,yurc2006}. External tether cutting or
``breakout'' reconnection is similar to tether cutting in that it is
a tension release mechanism via reconnection. But here it occurs
between the arcade envelope of the erupting field and an
over-arching restraining, reversed field of quadrupolar magnetic
configuration \cite{antioc1998,antioc1999}. Flux cancelation
\cite{martin1985,martin1989}, emergence of twisted flux ropes from
below the surface \cite{leka1996}, and ideal MHD instability
\cite{kliem2006,fan2007} are some other mechanisms proposed to
explain the eruption process.

Although there exist several proposed mechanisms, it is difficult to
disentangle  as to which particular mechanism is responsible for the
fast eruption in complex ARs. This difficulty arises due to the wide
variety of dynamic processes involved in such ARs. Therefore, we
have selected a filament eruption event that occurred in a
relatively simple solar AR NOAA 11093 on 7 August 2010 that led to
M1.0 class two-ribbon flare and a fast CME. Our aim in this study is
to understand the driver and trigger mechanisms of the eruption and
associated  processes. We investigate various possible conditions of
eruption process in relation to flux emergence/cancellation at
selected locations. From a morphological analysis, we attempt to
find triggering mechanism of the flare. We derive the flare
energetics in order to quantify flare characteristics and then
compare it with previous such studies.

The essence of high cadence and high resolution multi-wavelength
observations has already been revealed by earlier studies
\cite{liu2007,liu2005,wang1999}. In the present study, we utilize
the unique opportunity of coordinated observations in
multi-wavelength channels corresponding to different atmospheric
heights provided by Atmospheric Imaging Assembly (AIA;
\inlinecite{lemen2011}) and Helioseismic and Magnetic Imager  (HMI;
\inlinecite{schou2011}) on board Solar Dynamics Observatory(SDO).

The rest of the paper is organized as follows: The observational
data  and  reduction procedures are presented in Section~\ref{data}.
Results and discussions are described in Section~\ref{analys} while
Section~\ref{summ} gives the summary and conclusions.

%%%%%%%%%%%%%%%%%%%%%%%%%%%%%%%%%%%%%%%%%%%%%%%%%%%%%%%%%%%%%%%%%%%%%
%              OBSERVATIONAL DATA AND REDUCTION                     %
%%%%%%%%%%%%%%%%%%%%%%%%%%%%%%%%%%%%%%%%%%%%%%%%%%%%%%%%%%%%%%%%%%%%%

\section{Observational Data and Reduction}
\label{data}

The eruption event in AR NOAA 11093 ($N12^\circ E31^\circ$) occurred
on  7 August 2010. It produced a GOES  M1.0 class flare starting at
17:55 UT and peaking at 18:20 UT. This event was covered by SDO's
AIA and HMI, Reuven Ramaty High-Energy Solar Spectroscopic Imager
(RHESSI; \inlinecite{lin2002}), as well as, by the ground based GONG
H$\alpha$ network telescope at BBSO. The associated CME was detected
by the COR1 coronagraph \cite{thompson2003} on board  both the Ahead
and Behind satellites of the Solar TErrestrial RElations Observatory
(STEREO) which were separated by about $150^\circ$.

AIA takes multi-wavelength images at pixel size of
$0^{\prime\prime}.6$ pixel$^{-1}$    and  12 s cadence. To study the
flaring plasma, we focused on the images obtained in $\mbox{94\AA}$~
(Fe XVIII; $\log T = 6.8$), $\mbox{171\AA}$~  (Fe IX; $\log T =
5.8$) corresponding to the upper transition region, and
$\mbox{304\AA}$~ (He II; $\log T = 4.8$) corresponding to the
chromosphere and lower transition region. Images were added to
enhance the signal to noise ratio,  giving a cadence of 1 minute. We
have used preprocessed images (level 1.0) provided after
calibration, involving bad pixel correction, aligning, and scaling.

HMI makes measurements of line-of-sight magnetic field of the full
solar disk at $\mbox{6173\AA}$~ with a pixel size of
$0^{\prime\prime}.5$ and $45$ s cadence with a precision of 10G. We
rotated images (level 1.0) for solar north pointing up. Scaling of
data was done using the header information. We added every four
images for increasing the signal to noise ratio, giving a cadence of
3 minutes.

COR1 is an internally-occulted coronagraph and is  one of the STEREO
SECCHI suite of remote sensing telescopes. It takes observations of
CME from 1.3 to 4 solar radii in three different polarizing angles
every five minutes. We have used \texttt{secchi\_prep.pro} and
\texttt{cor1\_quickpol.pro} routines in STEREO software package to
process the images and finally to get total polarization brightness
images. For observing ejected material within 1.2 $R_\odot$, we have
also examined EUVI observations on board  STEREO-A.

We obtained H$\alpha$ 6563 \AA~ filtergrams at pixel size of
$1^{\prime\prime}$  and 1 minute cadence from the GONG telescope
operating at BBSO.

All full disk images obtained from different instruments were
aligned by differentially rotating to a reference image at 18:00 UT.
The offset was corrected after remapping and by overlaying magnetic
contours on images taken by different instruments. We have used
standard SolarSoftWare (SSW) library routines for our study.

%%%%%%%%%%%%%%%%%%%%%%%%%%%%%%%%%%%%%%%%%%%%%%%%%%%%%%%%%%%%%%%%%%%%%%%%
%                RESULTS AND DISCUSSIONS                               %
%%%%%%%%%%%%%%%%%%%%%%%%%%%%%%%%%%%%%%%%%%%%%%%%%%%%%%%%%%%%%%%%%%%%%%%%

\begin{figure}
\centering
\includegraphics[width=.9\textwidth, clip=,bb=43 27 416 228]{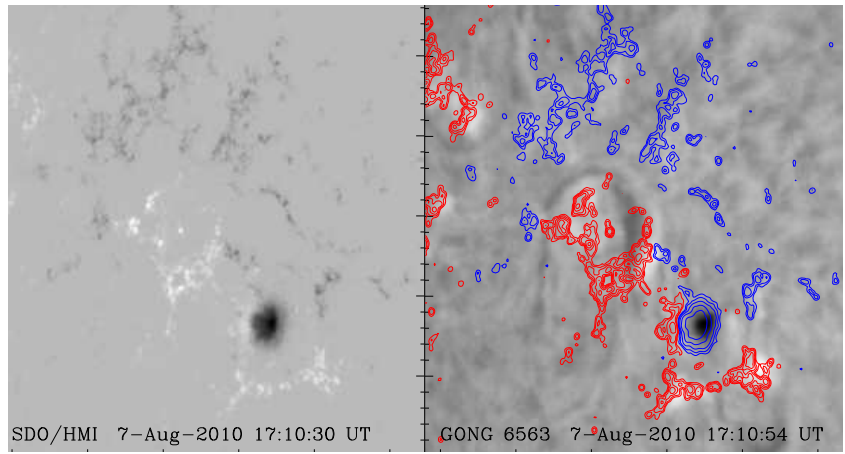}
\caption{HMI line of sight magnetogram (left panel), and GONG
H$\alpha$ filtergram (right panel) of NOAA 11093 overlaid with red
(blue) contours corresponding to positive (negative) magnetic
fluxes. North is directed upward in these and subsequent
maps.}\label{fig1}
\end{figure}

\section{Results and Discussions}
\label{analys}
\subsection{Filament Evolution Leading to the Two-Ribbon M1.0 Flare}

\subsubsection{Morphology}

\begin{figure}
\centering
\includegraphics[width=.75\textwidth,clip=,bb=6 14 425 490]{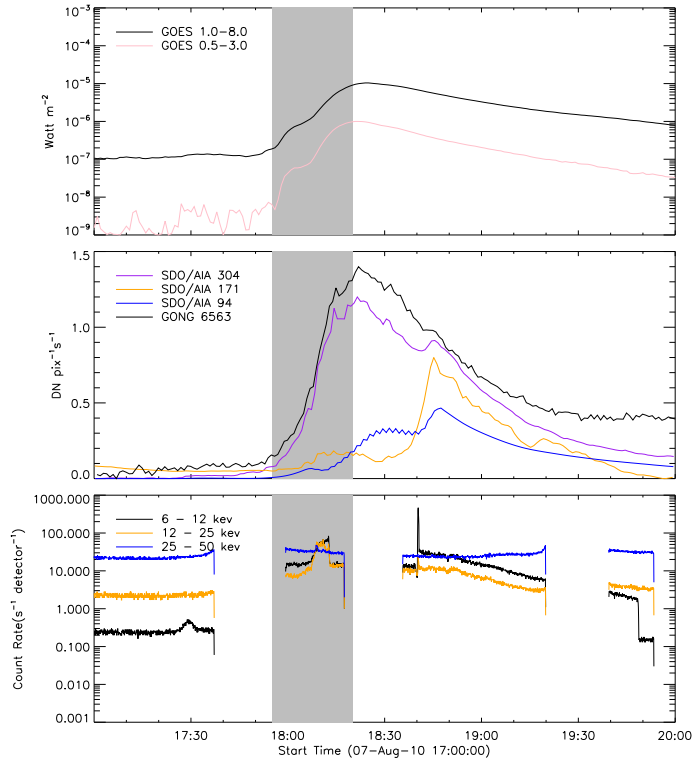}
\caption{Light curves of the  M1.0 class flare of 7 August 2010 in
AR NOAA 11093:   (top) GOES soft X-rays, (middle) AIA wavelengths
and GONG H$\alpha$, and (bottom) RHESSI hard X-rays. Gray shaded
region represents the impulsive phase of the flare as inferred from
GOES soft X-rays. Gaps in RHESSI light curves are due to the
spacecraft's passage through night-times and the South Atlantic
Anomaly.}\label{litcur}
\end{figure}

Figure~\ref{fig1}(left panel) shows an HMI magnetogram of AR NOAA
11093  taken on 7 August 2010 at 17:10 UT. The AR possessed a simple
magnetic configuration reported as $\beta-$ class by NOAA/USAF AR
Summary issued on 7 August 2010. The GONG H$\alpha$ filtergram of
the AR with contours of the line-of-sight magnetic field is shown in
the right panel. It consisted of a single main sunspot and an
inverse S-shaped filament with its one end connected to the sunspot,
as observed more than an hour before the eruption event ensued.
These images show the polarity of the dominant main sunspot and the
diffused fluxes of opposite polarities surrounding the filament.

The filament was oriented in nearly NE-SW direction  along the PIL.
Concerning the observed filament, the AR was bipolar with two main
leading polarities located on either side of PIL. We have
concentrated our study mainly on the time interval 17:00--20:00 UT
for identifying the changes occurring in morphological structure as
well as the connectivity of field-lines in the AR leading to
filament eruption and the flare.

Light curves of the flare in different wavelengths are shown in
Figure~\ref{litcur}. The gray shaded region from the start to peak
time of GOES flux represents the impulsive phase of the flare. The
decay phase of GOES soft X-rays flux lasted over three hours from
the peak time, implying that this relatively small M1.0 class flare
was a long duration event (LDE). The AIA 304 \AA~ and H$\alpha$
profiles essentially followed the GOES profiles, peaking at around
18:20 UT. However, AIA 94 and 171 \AA~ profiles did not agree well
in the impulsive phase, and peaked much later at 18:45 UT. By
examining the corresponding images, we found that the delayed
peaking in  these wavelengths corresponded with the post-flare loops
that appeared in the decay phase.

The RHESSI hard X-rays (HXR) profiles show two gaps owing to the
spacecraft's passage through the night-time and South Atlantic
Anomaly. However, the available data in the shaded region, i.e., the
impulsive phase of the flare, clearly shows a short duration
enhancement in the level of HXR emissions in 6-12 and 12-25 keV
channels. Notably, however, no enhancement was seen in the harder
emission in 25-50 keV (blue) channel.

\begin{figure}
\centering
\includegraphics[width=1.0\textwidth,clip=,bb=43 18 565 565]{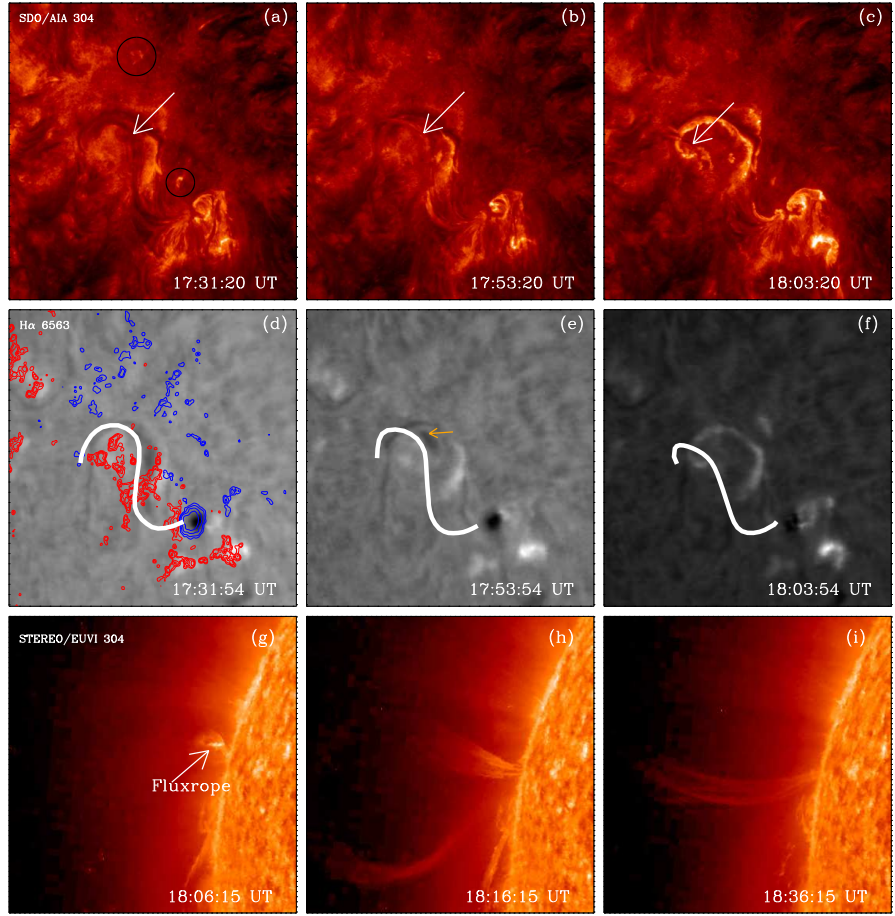}
\caption{(Top row) Images in AIA 304 \AA~ exhibit a rising motion of
flux rope lying along PIL (marked by arrows) with the brightening
region encircled. (Middle row) H$\alpha$ images showing
corresponding morphological changes in the chromosphere. White thick
curves drawn on these images represent the observed rising flux rope
as seen in 304 \AA~ images at respective times. In the frame (d),
overlaid red(blue) contours represent positive (negative) polarity
fluxes. (Bottom row) Flux rope rise observed in STEREO-A/EUVI 304
\AA~ at the east limb. }\label{preris}
\end{figure}

\begin{figure}[htbp]
\centering
\includegraphics[width=.98\textwidth,clip=,bb=24 14 420 220]{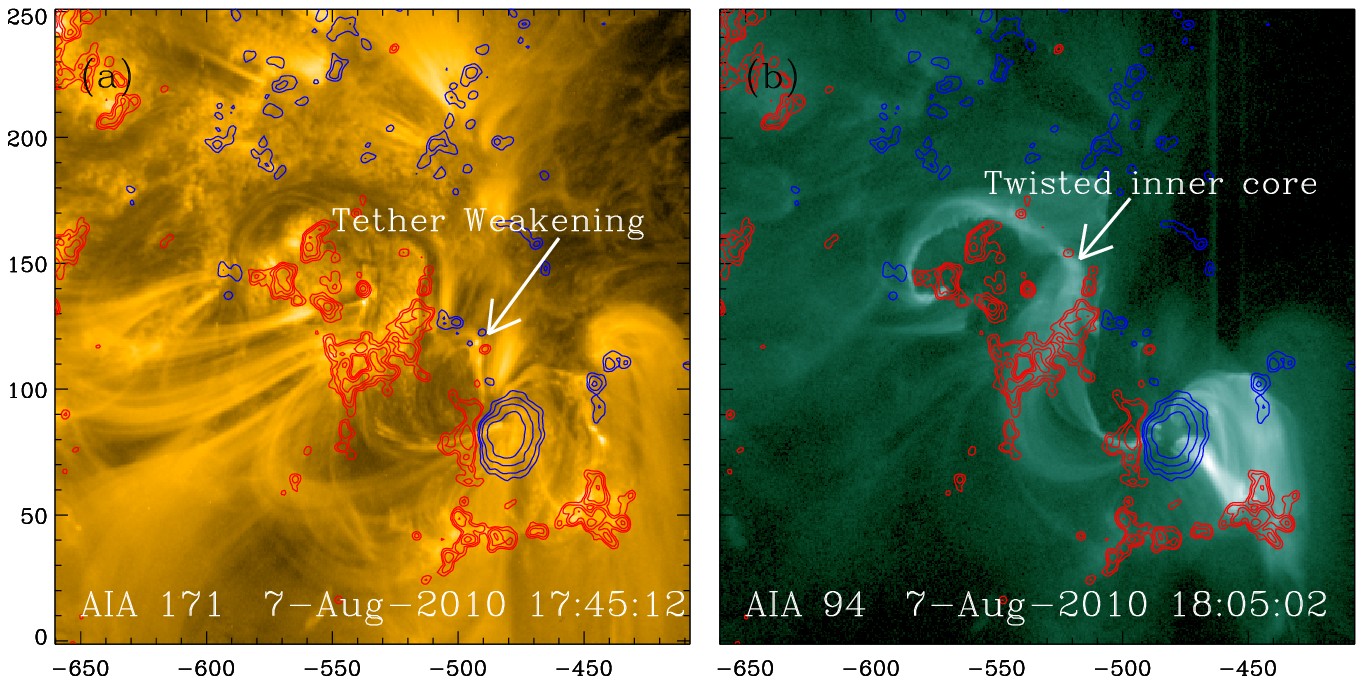}
\includegraphics[width=.98\textwidth,clip=,bb=0 0 456 218]{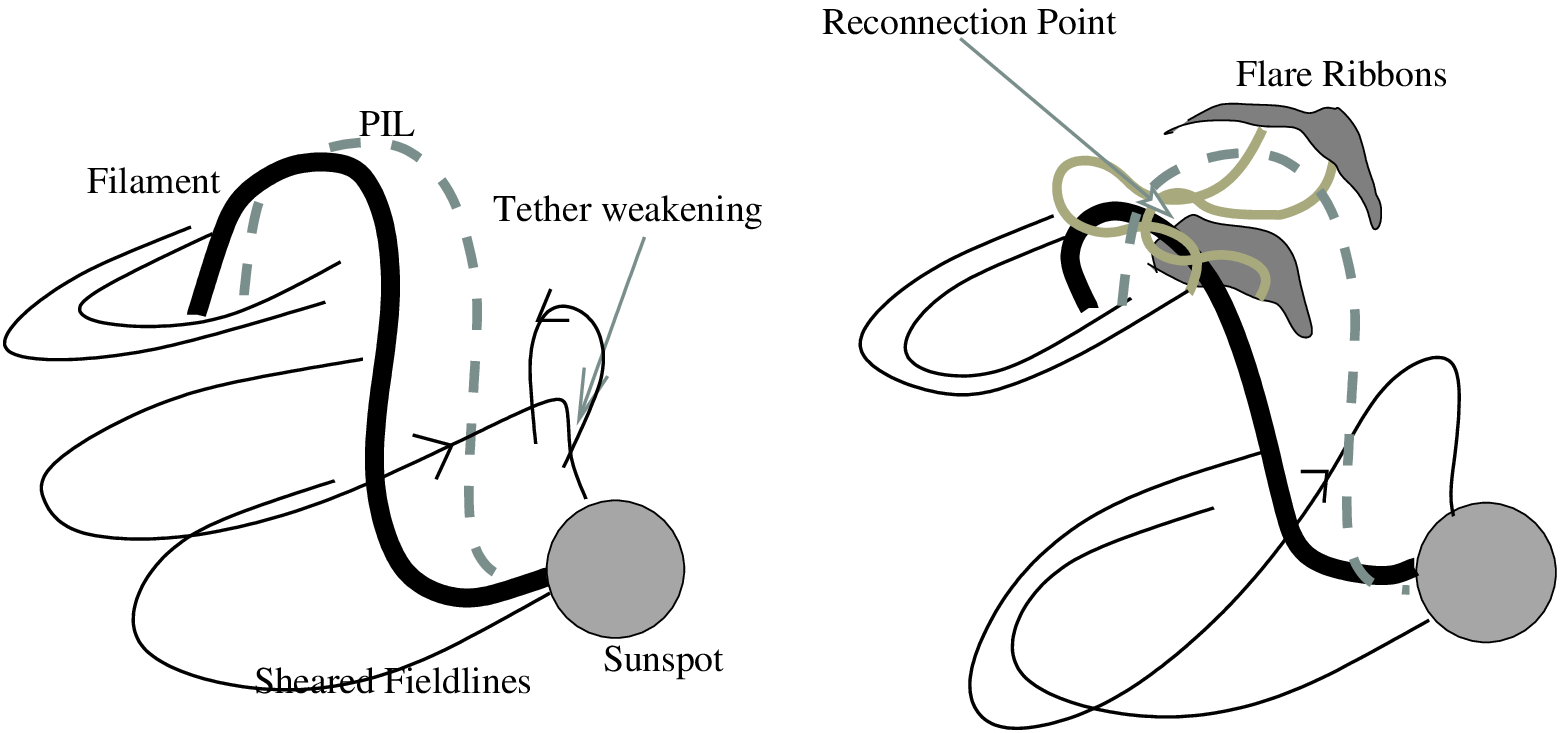}
\caption{Top: Field line topology before(a) and after(b) the onset
of eruption as seen in  94 and 171 \AA~ channels, respectively, with
superposed magnetic field contours. Bottom: Schematics of triggering
of flux rope/filament rise motion by tether weakening due to
interaction of side lobe field line with twisted overlying field
line from the sunspot. This rise motion enhanced the twist in the
inner core of the filament further, forming a current sheet beneath
it. Twisted inner core can be clearly seen in frame
(b).}\label{schem_fil_eru}
\end{figure}

>From movies made using the images in H$\alpha$ and AIA channels,  we
observed that  a flux rope having an average diameter of 6.9~Mm
($\sim9^{\prime\prime}.5$) started to rise from the inner core of
the filament at around 17:40 UT. Figure~\ref{preris}(a-c) shows, as
pointed by arrows, the rise of flux rope connecting the sunspot and
the other end of the filament elbow. The projection of this flux
rope on chromosphere is seen as a dark shadow moving toward east and
disappearing after 18:04 UT. At that time it had risen high, nearly
vertical to the disk plane. About 15 min after the start of its
rise, flare brightenings appeared; first the northern ribbon at the
hump part and then the southern one. The rise of filament was seen
as plasma jet ejection in EUVI images of STEREO A in  304 \AA~
(Figure~\ref{preris}(g-i)). STEREO A and B were separated by
150$^\circ$, and STEREO B was situated at 71$^\circ$ west from the
Earth, AR 11093 was 36$^\circ$ west of STEREO-B central meridian and
24$^\circ$ away from east limb of STEREO-A. Therefore, the flux rope
was not visible till 18:03 UT in STEREO-A. Furthermore, from
STEREO-B the event was not as clearly visible as in AIA 304 \AA~ due
to the lower spatial resolution (~1.6$^{\prime\prime}$/pixel).

We provide the schematics of eruption event constructed by a careful
examination of observed plasma tracers using animations of 171 and
94 \AA~ images in Figure~\ref{schem_fil_eru}. The EUV brightening in
the smaller encircled location of Figure~\ref{preris}(a) resulted
due to the interaction of overlying sheared field lines from the
sunspot with the side lobe field lines, as shown in
Figure~\ref{schem_fil_eru}(frame a and b). This interaction
substantially weakened the overlying field lines, allowing the flux
rope to rise upward. This rising motion formed current sheet in
cusped region of field lines connecting polarities on either side of
the filament leading to the onset of internal reconnection to
commence the flare. In 94 \AA~ image (frame b), the twisted bright
flux system can be seen in the inner core below the flux rope. It is
important to notice that the bright flux system appeared only in 94
\AA~ because soft X-ray emission began due to reconnection by the
above process. This scenario agrees well with the model of
\inlinecite{moore2001} that describes ``tether cutting'' as the
trigger mechanism. It is worth mentioning that the tether weakening
first occurred at the location of brightening and then induced
further in the core region beneath the filament. Tether cutting as a
trigger for the onset of eruption within the erupting system is
addressed by \inlinecite{yurc2006} in a quadrupolar configuration.
We will discuss the magnetic flux changes at the locations of these
brightenings in Section~\ref{FlxCha}

\begin{figure}
\centering
\includegraphics[width=1.0\textwidth,clip=,bb=35 10 560 560]{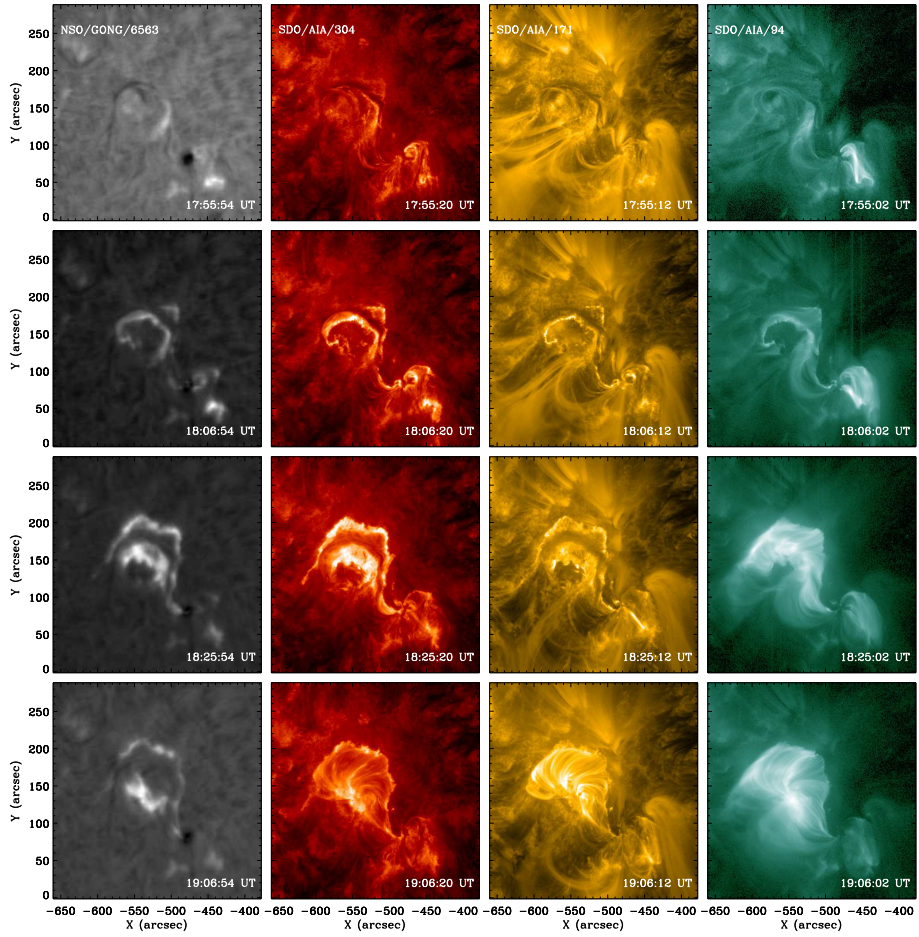}
\caption{A mosaic of images showing the filament evolution during
various  phases of  the flare -- start(row 1), impulsive(rows 2 and
3), and decay (row 4) as observed in GONG H$\alpha$ and AIA
wavelengths corresponding to successively higher atmospheric
layers.}\label{mosaic}
\end{figure}

\begin{figure}
\centering
\includegraphics[width=1.0\textwidth,clip=,bb=35 14 488 198]{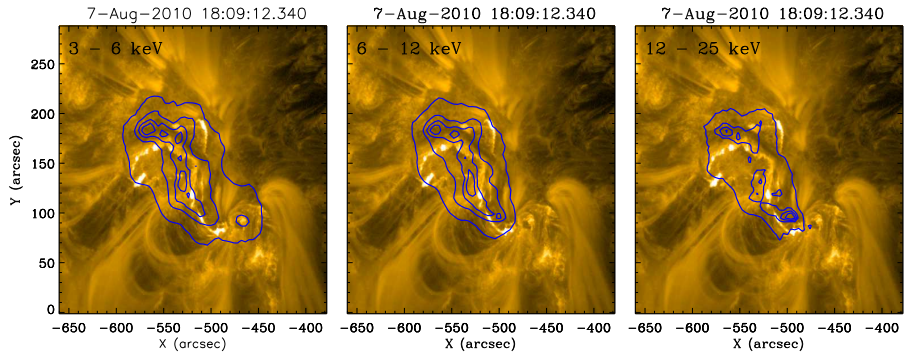}
\caption{Contours of RHESSI HXR overlaid on AIA 171 \AA~ image at
18:09 UT in  three energy bands: (left) 3-6 keV, (middle) 6-12 keV,
and (right) 12-25 keV, reconstructed by clean algorithm with one
minute integration time. Contour levels correspond to 50,70,80 and
90\% of respective peak flux.}\label{rheimg}
\end{figure}

Figure~\ref{mosaic} is a mosaic of multi-wavelength images
illustrating  filament eruption and the flare in H$\alpha$ and
various AIA channels corresponding to successively higher
atmospheric layers (from left to right columns). Rows correspond to
the time of start (top), impulsive (second and third), and decay
(bottom) phases of the flare. First column shows flare ribbons in
chromospheric H$\alpha$. As time progressed, flare ribbons
brightened and separated away up to the peak time 18:25 UT, and
decayed there after. Images in the second column correspond to AIA
304 \AA, showing flare ribbons in the upper chromosphere along with
the overlying field lines filled with $6\times10^4$K plasma.
Post-flare loops connecting the flare ribbons are clearly seen in
the decay phase. These post flare loops gave rise to the second peak
of 304 \AA~ light curve in Figure~\ref{litcur}. In third column, AIA
171 \AA~ images show plasma loops more clearly as these are
sensitive to $T \sim 6\times10^5$K. Fourth column corresponds to AIA
94 \AA~ images showing plasma loops at even higher temperatures
($\approx6\times10^6$K). We can observe here an increased twist in
core flux system due to tether cutting reconnection with increased
flux rope height (18:06:02 UT frame). It is important to note that
this twisted flux system is not visible in 304 and 171 \AA~
corresponding to lower heights. Once the main reconnection phase
commenced, these twisted flux system below the flux rope, as seen at
18:06:02 UT in the inner core of filament,  relaxed to arcades of
lower twist seen at 19:06:26 UT in the decay phase. This twisted, or
sigmoid to arcade evolution is an important mechanism of energy
release process as studied in many events \cite{liu2007,liu2005}. (A
movie, ``mosaic.mpeg'', is available on request.)

>From the RHESSI light curves, it is evident that HXR sources  were
produced during  the impulsive phase of the flare, at least in the
lower energies of 6-12 and 12-25 keV. Unfortunately no data was
available during 18:17-18:36 UT due to the spacecraft's night-time
and afterward when attenuators were out of the field of view. The
data with sufficient counts for imaging was available only during
18:00-18:10 UT in the impulsive phase of the flare.

We constructed the HXR images with ``clean'' algorithm from the
modulated  data and looked for HXR sources in the flaring region. In
Figure~\ref{rheimg}, we have plotted the contours of these
reconstructed images in the energy bands of 3-6, 6-12 and 12-25 keV
at 18:09 UT and overlaid on AIA 171 \AA~ images. Integration time of
the images was taken as 1 min; adequate to detect the changes. Due
to the rather low count rates in this event, it was not possible to
reconstruct images in  higher energy bands.

The HXR contours were found to be localized between the flare ribbon
kernels suggesting that reconnection took place at these locations,
where particles accelerated along the field lines and propagated
toward foot points anchored in the photosphere. In the 12-25 keV
image, a break in contours was observed on ribbon hump part. This
suggests that non-thermal HXR sources were located at foot points of
the flare ribbons. In the absence of additional HXR data, however,
it was not possible to follow up their further evolution.

\subsubsection{The Flare Energetics}
We can study the flare energetics by evaluating two main physical
parameters, viz.,  the rates of reconnection and energy release.
Reconnection rate is the electric-field strength in RCS and is
defined as the reconnected magnetic flux per unit time expressed as
\begin{equation}
\dot{\Phi} = B_{c} v_{in}.
\end{equation}
The magnetic energy release rate during a solar flare is the product
of Poynting  flux and the area of RCS that is generated during
magnetic reconnection. On the basis of reconnection model, it has
been shown by \inlinecite{isobe2002} that energy release rate can be
written as,
\begin{equation}
\frac{dE}{dt}=S A_{r}f_{r}=\frac{1}{2\pi}B_{c}^{2}v_{in}f_{r},
\end{equation}
where S is the Poynting flux into the reconnection region, $B_{c}$,
$v_{in}$, $A_{r}$ and $f_{r}$  are coronal magnetic field strength,
inflow velocity, area of the reconnection region, and the filling
factor of reconnection inflow, respectively.

The inflow velocity can be determined from observations. Using the
magnetic  flux conservation  theorem, one can write
\begin{equation}
B_{c} v_{in}=B_{\rm chro} v_{\rm ribb}=B_{\rm phot} v_{\rm ribb},
\end{equation}
where $v_{\rm ribb}$ is the velocity of the H$\alpha$ ribbon
separation,  and $B_{\rm phot}$,  $B_{\rm chro}$ are the
photospheric and chromospheric magnetic field strengths
respectively.

\begin{figure}
\centering
\includegraphics[width=.6\textwidth,clip=,bb=33 33 369 388]{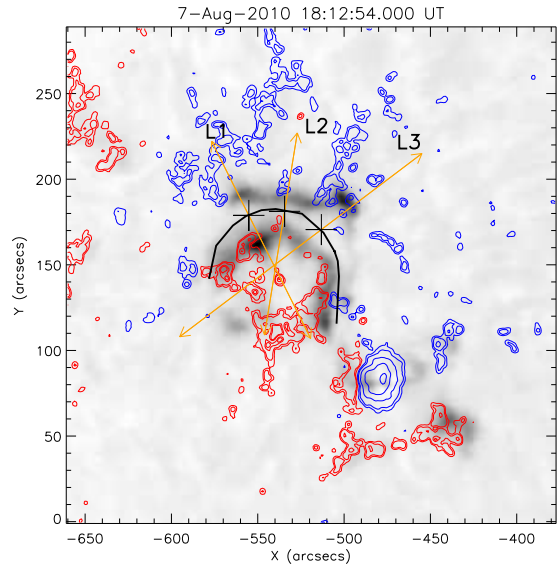}
\caption{H$\alpha$ image (in negative) of the flare  at
18:12:54\,UT, overlaid with magnetic  field contours. Solid black
line is the simplified PIL. Arrowed lines L1, L2, and L3 are drawn
perpendicular to PIL at points marked by ``+'' to follow the
separation of flare ribbons.}\label{nutlin}
\end{figure}

Figure~\ref{nutlin} depicts the scenario of flare ribbon separation
motion along the arbitrarily selected lines L1, L2, and L3. We have
measured distances from points marked as ``+'' on both sides of PIL
shown in thick solid line, i.e toward the north and south
directions. Separation velocities are then determined after fitting
Boltzmann sigmoids through the measurements. Further, using these we
calculated the reconnection rates, and Poynting fluxes for a
quantitative study of flare energetics.

\begin{figure}[htbp]
\begin{center}
\includegraphics[width=.7\textwidth,clip=,bb=8 8 314 497]{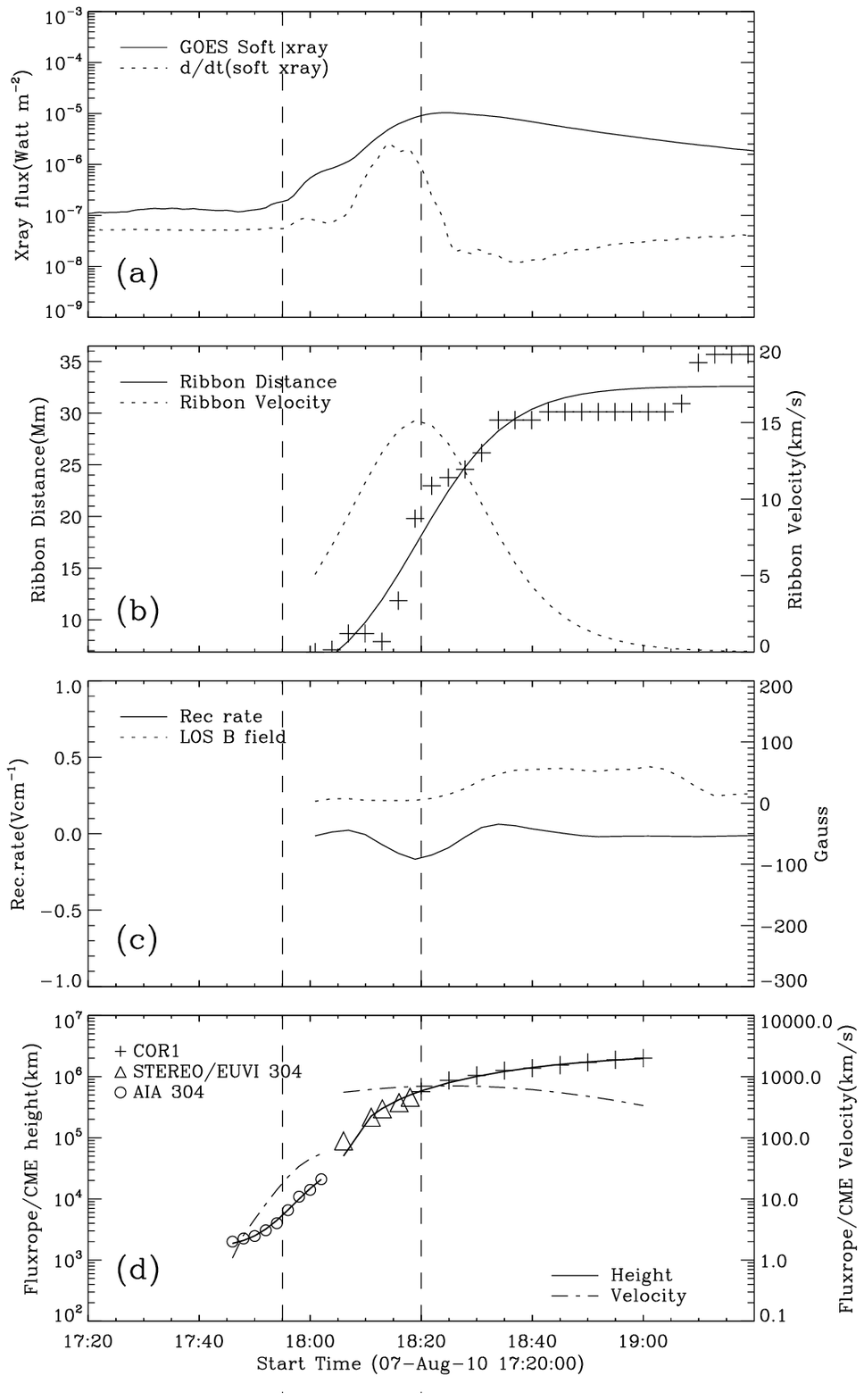}
\caption{(a) GOES X-ray flux and its derivative, (b) flare ribbons
separation distance and velocity profiles,  (c) reconnection rate
and Poynting flux,  and (d) filament/flux rope height as function of
time. Vertical dashed lines mark the start and peak times of flare
plotted for reference.}\label{flxrop_ht}
\end{center}
\end{figure}

For obtaining ribbons' separation velocity, we followed the
technique and  assumptions  as discussed in \inlinecite{maurya2010}
[and references therein] with $a = 0.2$ and $\epsilon=0.4$. We thus
obtained the reconnection rates in the range 0.5-3 Vcm$^{-1}$ and
Poynting fluxes in the range 0.01-3.8 G erg cm$^{-2}$s$^{-1}$ as
measured along the lines L1, L2, and L3. We have plotted one of the
temporal profiles in Figure~\ref{flxrop_ht} for comparing with the
filament/CME rise. This flare lasted for about 3000 seconds with
current sheet spread over an area $\approx 10^{19} cm^{2}$, with
average energy release rate of $10^{7}$ erg cm$^{-2}$s$^{-1}$. Using
these typical values, we estimate the total energy released during
the flare event to be a modest $10^{29} $ergs, in conformity with
the magnitude of the flare. Our results are consistent with that of
quiescent filament eruption reported by \inlinecite{wang2003}.

We notice here that kernel velocities were not uniform along the
selected lines. Further, electric field that is expected to release
energy through dissipation of electric currents in RCS, temporally
did not correlate well with impulsive hard X-ray emission in some
directions. To investigate the electron acceleration in the
impulsive phase of the flare, we require hard X-ray and microwave
observations. But, due to the data gap in hard X-rays during the
impulsive phase, we have used the time derivative of GOES soft X-ray
light curve as an alternative to the hard X-ray with the assumption
that Neupert effect is valid in this event (as in reference XXXX).

Our estimates for this M1.0 are lower as compared to the velocities
of 15-50 kms$^{-1}$ and reconnection rates of 2.7-11.8 Vcm$^{-1}$
 for the M3.9 flare event reported by \inlinecite{miklenic2007}. Also, for
a rather short duration, small C9.0 flare, \inlinecite{qiu2002}
reported a significantly larger kernel velocity of 20-100 kms$^{-1}$
and peak electric field of 90 Vcm$^{-1}$. They found ribbon motions
both parallel and perpendicular to PIL and concluded that either the
2D-magnetic reconnection theory related to the H$\alpha$ kernel
motion was applicable only to a part of the flare region due to its
particular magnetic geometry, or the electron acceleration was
dominated by some other mechanisms depending on the electron energy.
We therefore suggest that the estimated electric fields and
reconnection rates depend not only on the magnitude of the flare,
but also on the flare kernel velocity and the magnetic field
geometry.

\subsection{The CME and its dynamics}
\label{erucme} The  flux rope started rising at 17:40 UT as seen in
AIA 304\AA~  (cf., Figure~\ref{preris}). Further, with its rise, the
surrounding overlying  loops also started rising as if they formed a
balloon like cavity expanding temporally. The jet like ejection of
plasma came out from the reconnection region with ejection velocity
presumably proportional to the reconnection rate as its height
increased. Eventually,  it is observed as the CME event in COR1
coronagraph at about 18:20 UT.

We project the lateral displacements of filament in vertical
direction in  AIA 304 \AA~ in pre-rise phase(see for details
\inlinecite{wang2003}) and corrected heights of flux rope as it is
seen exactly on the limb in EUVI/STEREO in impulsive phase. Then we
reconstructed the heights of CME front end from triangulation method
in decay phase to obtain the heights of filament/flux with its rise.
%Note that these measurements involve subjective errors and
%repetition can not reproduce the same numbers.
In order to reduce the measurement errors, we fit them with Boltmann
sigmoid function as used for flare ribbon motion \cite{maurya2010}.
This function suits well as a model to the data because the lower,
steep and upper parts of this function resemble the rise, impulsive
and decay phases of CME/flux rope dynamics and can be fitted by four
parameters in the least square sense.

Heights of filament and CME front is plotted in
Figure~\ref{flxrop_ht} for comparing with temporal evolution of
flare ribbon separation, filament eruption and flare energy release
in terms of reconnection rate \cite{wang2003}. The fast-rising stage
coincided with the flare impulsive phase, and the mass acceleration
increased rapidly along with the increase of magnetic reconnection
rate. As evident from the figure, flare ribbons appeared at 17:58 UT
and attained maximum velocity at the peak phase of flare. This
motion agreed with hard X-ray profile as electrons injected onto
photosphere in the impulsive phase with increasing reconnection
rate. From the figure, we can see that filament started to rise in
the first 15 minutes with velocity in the range 8-10 kms$^{-1}$
reaching 100 kms$^{-1}$ at an average acceleration of 60 ms$^{-2}$.
>From STEREO observations including COR1, we found that the CME
traveled with an average speed of 590 kms$^{-1}$ and reached peak
acceleration of 220 ms$^{-2}$ at the peak phase of the flare. It
then decelerated in the decay phase gradually. This is in
correspondence with magnetic reconnection rate and flux rope
acceleration obtained from a study of 13 well observed two-ribbon
flares \cite{jing2005}. The acceleration in the range 50-400
ms$^{-2}$ and peak reconnection rates in the range 0.2-5.0
Vcm$^{-1}$ as obtained by them are consistent with our results.

\subsection{Changes in the Photospheric Magnetic Field }
\label{FlxCha} Theoretical models suggest that evolution of magnetic
fields at or below the photosphere,  in the form of flux emergence
and cancellation, could result in a loss of equilibrium of the
magnetic structures \cite{martin1985,martin1989,leka1996}. Changes
in the photospheric longitudinal magnetic field around the time of
eruption have been examined by many workers in the past
\cite{wang1999,mathew2000,green2003,ambastha2007,jiang2007,sterl2007}.
\inlinecite{zhang2008b} studied the relationship between flux
emergence and CME initiation inferring that $60\%$ of CME source
regions have increase and $40\%$ have decrease of magnetic flux.

Here, we look for regions, if any, of flux emergence/cancellation in
photospheric LOS magnetic field using high resolution SDO/HMI
magnetograms. We carried out registration of the images by
differentially rotating to a reference image at 18:00 UT. Effects of
telescope jitter and other pointing errors were corrected by using a
cross correlation method reducing the uncertainty within 1-2 arc
sec. Every four images were added to yield a cadence of 3min. As the
HMI precision is 10G, we neglected magnetic fields below this value.

\begin{figure}[htbp]
\centering
\includegraphics[width=.50\textwidth,clip=,bb=82 393 362 670]{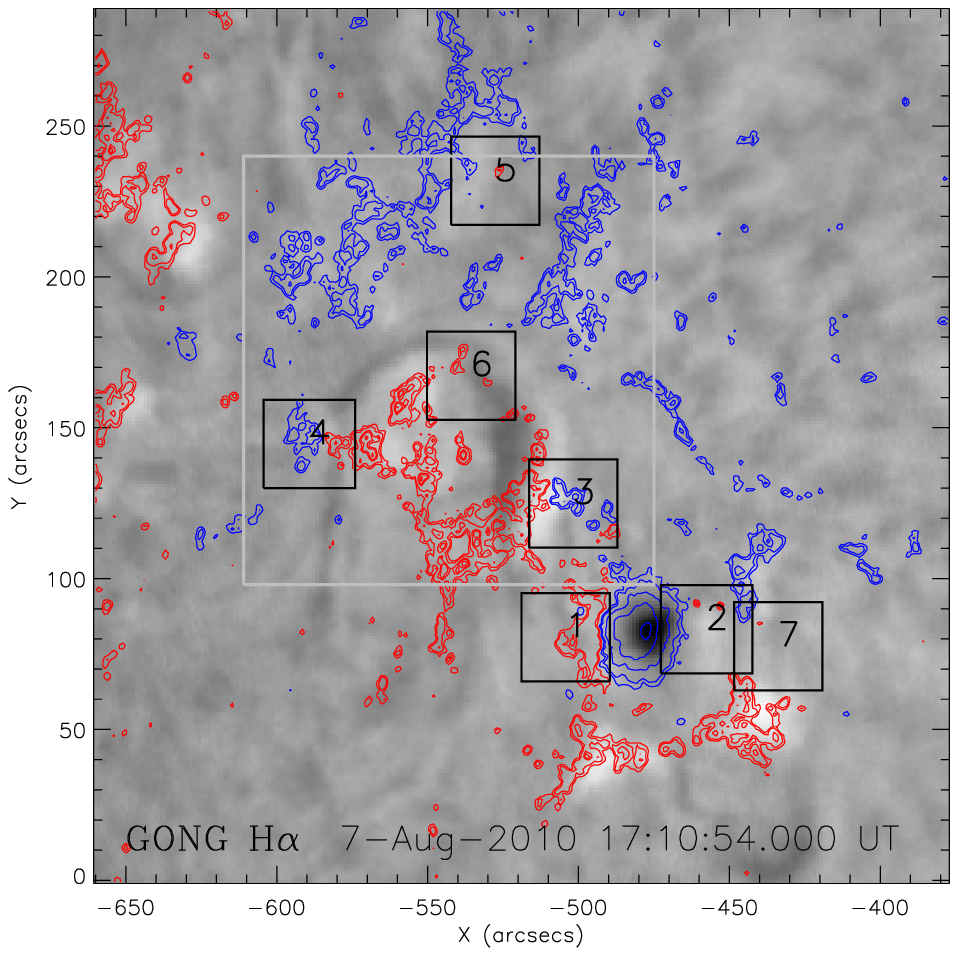}
\includegraphics[width=.48\textwidth,clip=,bb=28 8 356 346]{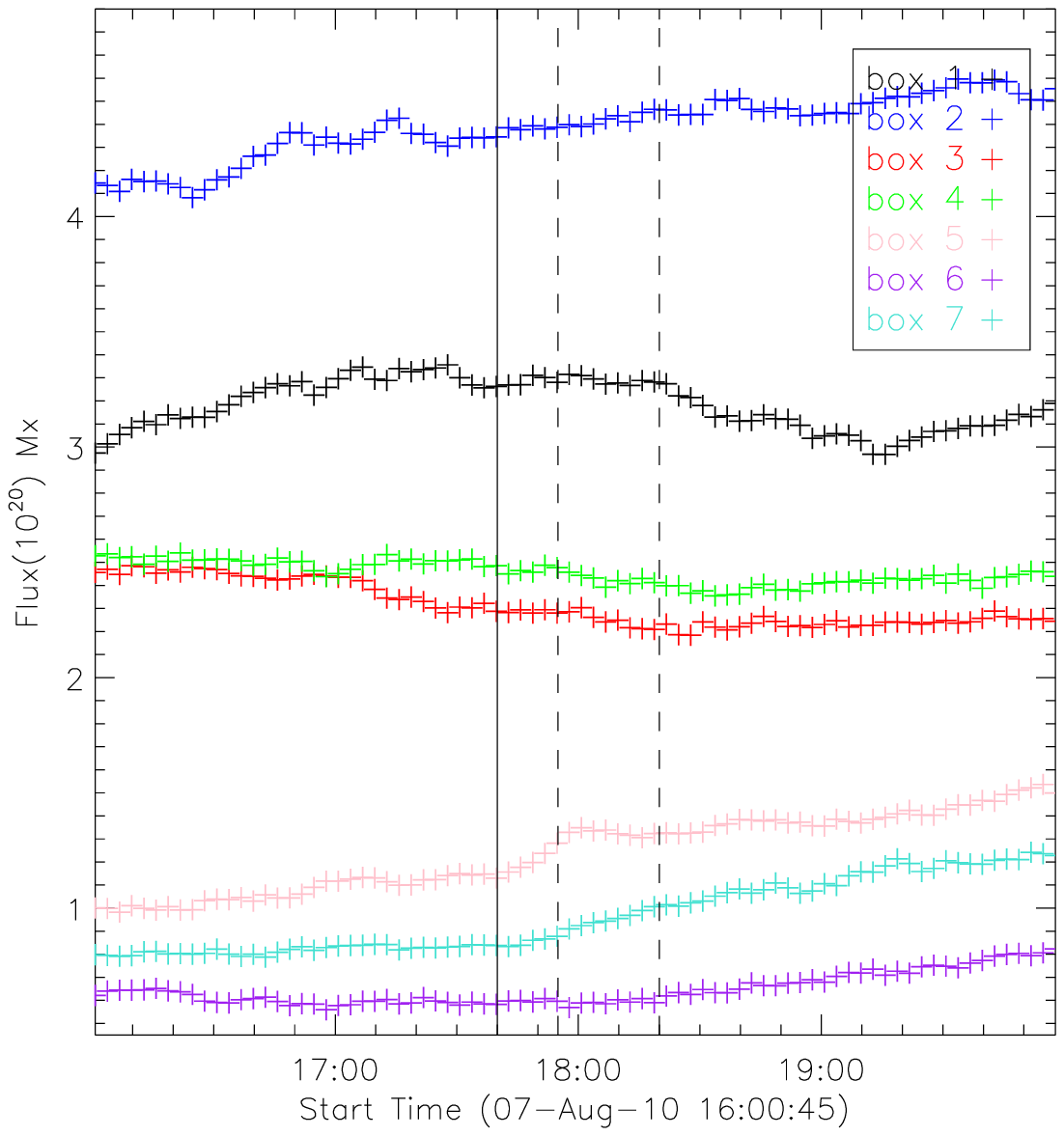}
\caption{(Left) GONG H$\alpha$ image of NOAA11093 overlaid with HMI
LOS magnetogram contours. Boxes(1-7) mark the selected
regions-of-interest (ROI) for calculation of unsigned flux.
(Right):~Temporal profiles of unsigned flux in the
ROIs.}\label{magbx_prof}
\end{figure}

\begin{figure}[htbp]
\begin{center}
\includegraphics[width=.95\textwidth,clip=,bb=12 8 358 274]{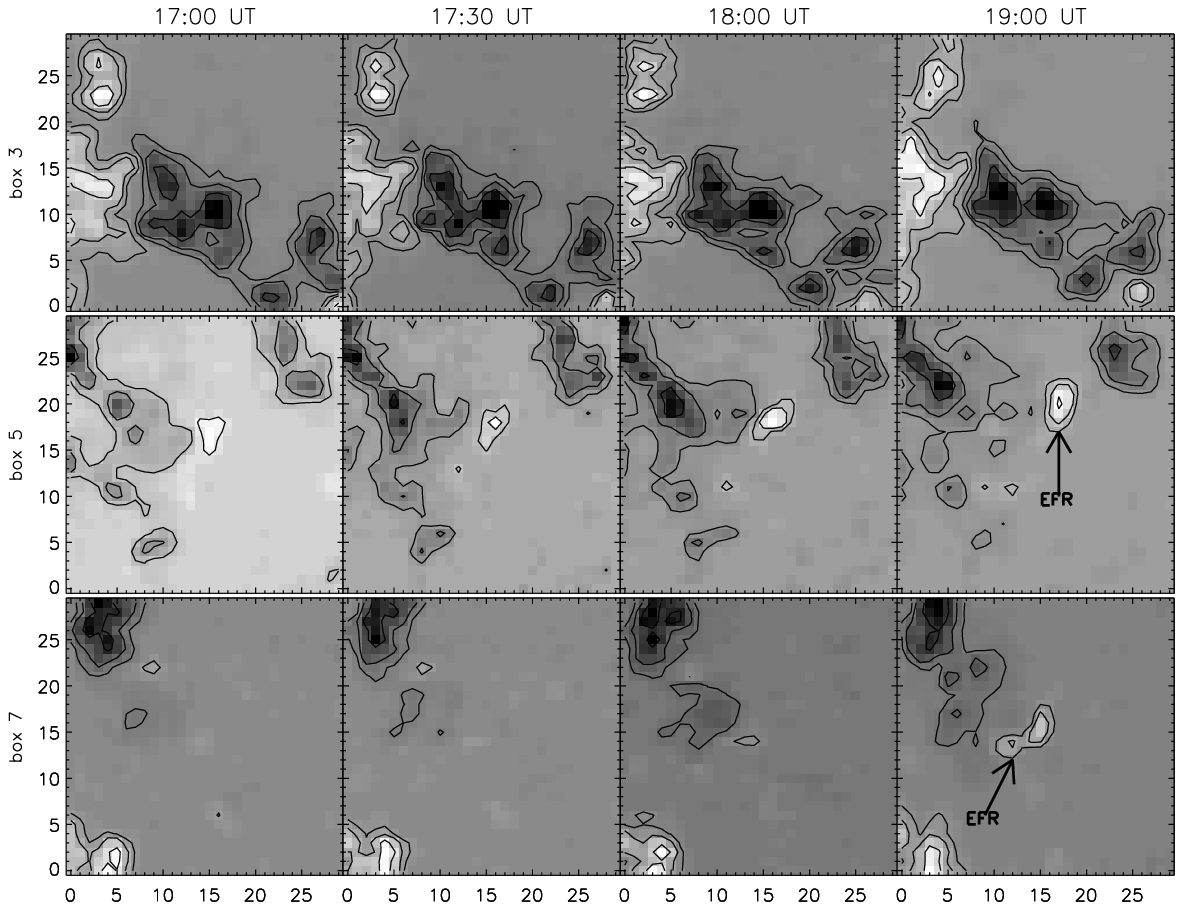}
\includegraphics[width=.95\textwidth,clip=,bb=17 48 358 234]{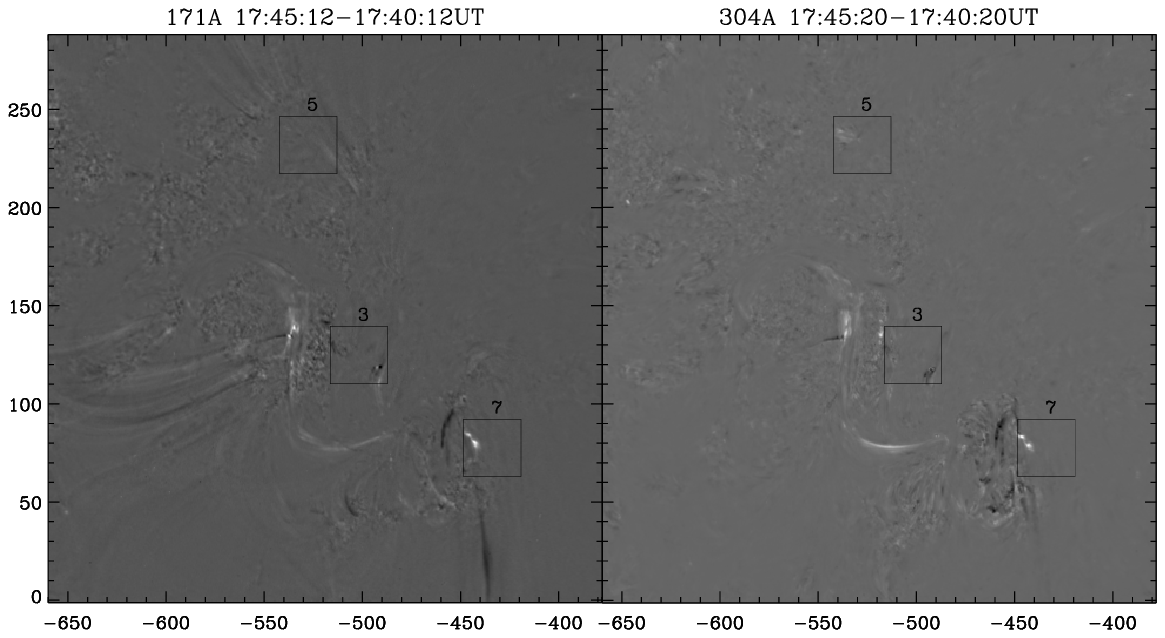}
\caption{Enlarged box regions 3,5 and 7 of Figure~\ref{magbx_prof},
showing magnetic flux evolution with time (top panel). Difference
images in pre-eruption/flare phase of 17:45 - 17:40 UT in AIA 171
\AA~, 304 \AA~ wavelengths showing associated brightenings (bottom
panel).}\label{brig}
\end{center}
\end{figure}

A typical GONG H$\alpha$ image of NOAA 11093 overlaid with contours
of LOS magnetic fields is shown in Figure~\ref{magbx_prof} (left
frame). From a movie of the registered images, we identified sites
of flux emergence/cancellation, marked in the figure by boxes,
located around the filament and the sunspot in the AR. Time profiles
of unsigned magnetic fluxes corresponding to the selected boxes are
plotted during the period 16:00-20:00 UT. Sufficient care was taken
in selecting the box size of
$30^{\prime\prime}\times30^{\prime\prime}$. A very small box size
would not adequately cover the region of interest, while averaging
over too large a region would dilute the magnitude of changes.

We interpret temporal evolution of fluxes in each box and its
contribution to the  stability of sunspot-filament magnetic system.
Boxes 1 and 2 are located around the leading sunspot. Box 1 selected
at the penumbral location of sunspot also covered regions of
opposite polarity fluxes around the filament. Total unsigned flux
increased there till the time of onset of the filament eruption, and
decreased thereafter. On the other hand, changes were oscillatory in
box 2 before the flare, where a small region of negative (red) flux
emerged along with pre-flare brightening.

Boxes 3, 4, 5 and 6 are located at either side of the PIL from where
field lines  originated to tie the filament as ropes required for a
stable configuration. Changes in these boxes are important for
examining the stability of the filament. In box 3, a gradual
decrease of $0.2\times10^{20}$ Mx flux occurred at the flare onset
time from the time of start of the rope rise. On the other hand,
flux increased in boxes 5, 6, and 7 as a result of new flux
emergence. Of these, box 6 was located under the flux rope.
Evidently, there were sites of flux emergence/cancellation in and
around the filament influencing its stability. As the HMI
measurements do not suffer by the degrading effect of Earth's
atmosphere, one can be reasonably confident about the observed flux
changes.

To further corroborate these temporal changes, we enlarged some ROIs
as shown in Figure~\ref{brig} (top panel). Emerging flux
regions(EFR), marked by arrows, can be seen in these regions.
However, we did not find appreciable disappearance or cancelation in
box 3. We also looked for any brightening expected to be associated
with flux emergence/cancellation, by examining difference images of
pre-eruption/flare phase, i.e., 17:45 -17:40 UT, in AIA 171 \AA~ and
304 \AA~ (Figure~\ref{brig} (bottom panel)). This showed a
brightening (seen as darkening in the difference images) in box 3
where a gradual decrease of unsigned flux was observed during the
rise phase of the filament. Similar changes were also found in boxes
5 and 7 (seen as white dots in difference images).

\begin{figure}
\centering
\includegraphics[width=.5\textwidth,clip=,bb=9 1 280 352]{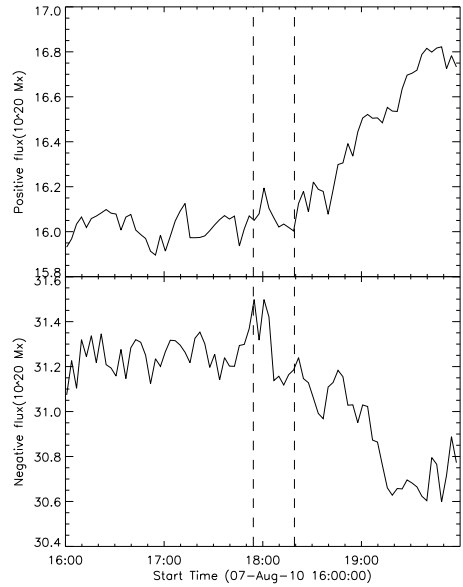}
\caption{Time profiles of positive and negative flux within the
region  covering the entire filament area (gray big box in
Figure~\ref{magbx_prof}(left pannel)). The start and peak times of
the flare are marked by the dashed vertical lines.}\label{flx_chan}
\end{figure}

In a recent work, \inlinecite{wang2010} found observational evidence
of back reaction on the solar surface fluxes with coronal magnetic
restructuring after reconnection in well observed X-class flares.
They suggested that such results may also be detectable for smaller
flares from the low threshold HMI data. To investigate it, we
computed positive and negative fluxes separately in the region shown
as the large box (drawn in gray color) covering the filament
(Figure~\ref{magbx_prof}(left panel)). Figure~\ref{flx_chan} shows
that positive/negative fluxes increased/decreased by similar amount
of $\approx 10^{20}$ Mx in the post flare phase. This gives an
indirect evidence of reconnection as a consequence of tether
cutting, mentioned in the previous sections.

Thus, we have found regions of flux emergence, cancelation and
associated EUV brightening in some locations of the AR (see
Figure~\ref{preris}(a)). The interaction of field lines at one of
these brightening regions gives an evidence of reconnection leading
to tether weakening. These processes could have destabilized the
filament to rise upward leading to tether cutting reconnection. The
observed photospheric flux changes found in the decay phase of the
event are in conformity with \inlinecite{wang2010}, i.e., evidence
of the back reaction on solar surface as a consequence of
reconnection.

%%%%%%%%%%%%%%%%%%%%%%%%%%%%%%%%%%%%%%%%%%%%%%%%%%%%%%%%%%%%%%%%%%%%%%%%%%%
%                 SUMMARY AND DISCUSSION                                  %
%%%%%%%%%%%%%%%%%%%%%%%%%%%%%%%%%%%%%%%%%%%%%%%%%%%%%%%%%%%%%%%%%%%%%%%%%%%
\section{Summary and Conclusions}
\label{summ}

The current models assume that the stored energy lies in a low lying
magnetic flux system which is twisted or sheared. This system is the
so-called core flux or flux rope. Such twisted systems in the form
of inverse S (or forward S shape), called sigmoid, are usually seen
in association with eruptions as the present event.
\inlinecite{canf1999} had studied the nature of activity with
respect to morphology of eruptive or non-eruptive events with soft
X-ray observations and inferred that sigmoids are more likely to
erupt.

In this paper, we have presented multi-instrument multi-wavelength
observations and analysis of the eruption of an inverse S shaped
filament. This event exhibits a good example of standard solar flare
characterized by the filament eruption, two ribbon separation and
its association with a fast CME. We summarize below the main
findings of our analysis of this eruption event:

\begin{enumerate}
  \item From morphological study of this event, we
  inferred that the rising motion of the filament was triggered by
  remote tether weakening at coronal brightening region,
  which further induced tether cutting reconnection
  underneath it to unleash the eruption process, leading to a fast CME
  subsequent to the two-ribbon flare.

  \item Flare ribbons or kernels separated out with modest velocities in the range
   12-16~kms$^{-1}$. This was used to estimate various physical parameters for
   evaluating the flare energetics using a 2-D model. Reconnection rates
   and Poynting fluxes were estimated in the range of 0.5-3.0Vcm$^{-1}$ and
  0.01-3.8 G erg cm$^{-2}$s$^{-1}$, respectively. These were sufficient
  to release free energy of $10^{29}$ ergs of an M-class flare.

   \item Filament/flux rope rising motion profile indeed showed
  correspondence with various flare characteristics, viz., reconnection rate and
  hard X-ray emission profiles. It started rising with velocity of 8-10
  kms$^{-1}$, reaching a maximum of around 100 kms$^{-1}$ with
  an average acceleration of 60 ms$^{-2}$ (estimated by projecting lateral
  displacements of filament on to the vertical
  direction observed in 304 \AA~ channel of AIA). Further, this flux rope accelerated
  to the maximum velocity as the CME, observed at the peak phase of
  the flare, followed by its deceleration to an average velocity of 590 kms$^{-1}$.

  \item Flux variations in and around the filament were examined
  before and after the eruption. We found some areas of flux changes,
  co-temporal with the onset of
  filament rise. For example, gradual reduction of positive flux was
  found in box 3 (cf., Figure~\ref{magbx_prof}) at the onset time of filament rise.
  From a careful study of 171 \AA~ images, it is interpreted as a consequence
  of interaction of overlying field lines across the filament with side lobe field
  lines resulting in tether weakening of the sigmoidal filament
  system. In addition, flux emergence in box 5 located near the rising part of
  the filament might have contributed to destabilize the system.
  In summary, we infer that these flux changes
  caused the loss of equilibrium leading to slow, upward rise of the
  filament, and the onset of eruption by tether cutting reconnection.
  In turn, changes occurred in photospheric fluxes in the decay
  phase of the flare as a back reaction of this
  reconnection (Figure~\ref{flx_chan}), in accordance with the
  recent findings of \inlinecite{wang2010}.

\end{enumerate}

Destabilization of the filament system can occur due to either
ideal-MHD, or kink,  instability \cite{kliem2006}. An increased
twist in the filament or flux rope system can become kinked by flux
emergence, because of which filament itself can rise. Another
possibility could be shear motions of photospheric fluxes; not
studied here. We suggest that the observed emergence/cancellation of
magnetic fluxes near the filament caused the flux rope to rise,
resulting in the tethers to cut and reconnection to take place
beneath the filament; in agreement with the tether cutting model. We
intend to pursue this study further by invoking observations of
vector magnetograms as boundary conditions for extrapolations to
look for changes in the coronal magnetic field and other associated
parameters.

%%%%%%%%%%%%%%%%%%%%%%%%%%%%%%%%%%%%%%%%%%%%%%%%%%%%%%%%%%%%%%%%
%                   ACKNOWLEDGEMENTS
%%%%%%%%%%%%%%%%%%%%%%%%%%%%%%%%%%%%%%%%%%%%%%%%%%%%%%%%%%%%%%%
\begin{acks}
The AIA(HMI) data used here are courtesy of SDO(NASA) and the
AIA(HMI)  consortium. We  thank AIA team for making available the
processed data. This work utilizes data obtained by the Global
Oscillation Network Group (GONG) Program, managed by the National
Solar Observatory, which is operated by AURA, Inc., under a
cooperative agreement with the National Science Foundation. We thank
the anonymous referee for carefully going through the manuscript and
making valuable comments which improves readability of manuscript
appreciably.
\end{acks}

%%%%%%%%%%%%%%%%%%%%%%%%%%%%%%%%%%%%%%%%%%%%%%%%%%%%%%%%%%%%%%
%                    BIBLIOGRAPHY
%%%%%%%%%%%%%%%%%%%%%%%%%%%%%%%%%%%%%%%%%%%%%%%%%%%%%%%%%%%%%%

%\bibliographystyle{spr-mp-sola-cnd}
%\bibliography{ref_bulk}

\end{article}
\end{document}